\documentclass{sig-alternate-05-2015}

\usepackage{graphicx}
\usepackage[usenames, dvipsnames]{color}

\begin{document}

% Copyright
%\setcopyright{acmcopyright}
%\setcopyright{acmlicensed}
%\setcopyright{rightsretained}
%\setcopyright{usgov}
%\setcopyright{usgovmixed}
%\setcopyright{cagov}
%\setcopyright{cagovmixed}

% DOI
%\doi{10.475/123_4}

% ISBN
%\isbn{123-4567-24-567/08/06}

%Conference
%\conferenceinfo{PLDI '13}{June 16--19, 2013, Seattle, WA, USA}

%\acmPrice{\$15.00}

%
% --- Author Metadata here ---
%\conferenceinfo{WOODSTOCK}{'97 El Paso, Texas USA}
%\CopyrightYear{2007} % Allows default copyright year (20XX) to be over-ridden - IF NEED BE.
%\crdata{0-12345-67-8/90/01}  % Allows default copyright data (0-89791-88-6/97/05) to be over-ridden - IF NEED BE.
% --- End of Author Metadata ---

\title{Multi-dimensional Features for Prediction with Tweets}
%\subtitle{[Extended Abstract]
%\titlenote{A full version of this paper is available as
%\textit{Author's Guide to Preparing ACM SIG Proceedings Using
%\LaTeX$2_\epsilon$\ and BibTeX} at
%\texttt{www.acm.org/eaddress.htm}}}
%
% You need the command \numberofauthors to handle the 'placement
% and alignment' of the authors beneath the title.
%
% For aesthetic reasons, we recommend 'three authors at a time'
% i.e. three 'name/affiliation blocks' be placed beneath the title.
%
% NOTE: You are NOT restricted in how many 'rows' of
% "name/affiliations" may appear. We just ask that you restrict
% the number of 'columns' to three.
%
% Because of the available 'opening page real-estate'
% we ask you to refrain from putting more than six authors
% (two rows with three columns) beneath the article title.
% More than six makes the first-page appear very cluttered indeed.
%
% Use the \alignauthor commands to handle the names
% and affiliations for an 'aesthetic maximum' of six authors.
% Add names, affiliations, addresses for
% the seventh etc. author(s) as the argument for the
% \additionalauthors command.
% These 'additional authors' will be output/set for you
% without further effort on your part as the last section in
% the body of your article BEFORE References or any Appendices.

\numberofauthors{2} %  in this sample file, there are a *total*
% of EIGHT authors. SIX appear on the 'first-page' (for formatting
% reasons) and the remaining two appear in the \additionalauthors section.
%
\author{
% You can go ahead and credit any number of authors here,
% e.g. one 'row of three' or two rows (consisting of one row of three
% and a second row of one, two or three).
%
% The command \alignauthor (no curly braces needed) should
% precede each author name, affiliation/snail-mail address and
% e-mail address. Additionally, tag each line of
% affiliation/address with \affaddr, and tag the
% e-mail address with \email.
%
% 1st. author
\alignauthor 
Nupoor Gandhi, Alex Morales\\
       \affaddr{Department of Computer Science}\\
       \affaddr{University of Illinois at Urbana-Champaign}\\
       \email{{nupoorg2, amorale4}@illinois.edu}
\alignauthor
Dolores Albarracin\\
       \affaddr{Department of Psychology}\\
       \affaddr{University of Illinois at Urbana-Champaign}\\
       \email{dalbarra@illinois.edu}
  % use '\and' if you need 'another row' of author names
% 4th. author
% 5th. author
%\alignauthor Sean Fogarty\\
% 6th. author
%\alignauthor Charles Palmer\\
}
% There's nothing stopping you putting the seventh, eighth, etc.
% author on the opening page (as the 'third row') but we ask,
% for aesthetic reasons that you place these 'additional authors'
% in the \additional authors block, viz.
% \additionalauthors{Additional authors: John Smith (The Th{\o}rv{\"a}ld Group,
% email: {\texttt{jsmith@affiliation.org}}) and Julius P.~Kumquat
% (The Kumquat Consortium, email: {\texttt{jpkumquat@consortium.net}}).}
% \date{30 July 1999}
% Just remember to make sure that the TOTAL number of authors
% is the number that will appear on the first page PLUS the
% number that will appear in the \additionalauthors section.

\maketitle
%\begin{abstract}
% abstract is optional for this format
%\end{abstract}

\section{Introduction}

With the rise of opioid abuse in the US, there has been a growth of overlapping hotspots for overdose-related and HIV-related deaths in Springfield, Boston, Fall River, New Bedford, and parts of Cape Cod \cite{opioidwebsite}. With a large part of population, including rural communities, active on social media, it is crucial that we leverage the predictive power of social media as a preventive measure.\newline
\indent We explore the predictive power of micro-blogging social media website Twitter with respect to HIV new diagnosis rates per county. 
While trending work in Twitter NLP has focused on primarily text-based features, we show that multi-dimensional feature construction  can significantly improve the predictive power of topic features alone with respect STI's (sexually transmitted infections). By multi-dimensional features, we mean leveraging not only the topical features (text) of a corpus, but also location-based information (counties) about the tweets in feature-construction. We develop novel text-location-based smoothing features to predict new diagnoses of HIV.
\newline
\indent Previous predictive work with Twitter data has leveraged part of speech and semantic structure to build topics.
There has been extensive work in the analysis of sentence structure and language semantics.\cite{politicalstatements} forms a relation between individuals and their opinions to learn latent positions of people and propositions. \newline
\indent Despite the 140-character limit on tweets, there have been numerous studies with Twitter where  semantic features produced strong results. For example, \cite{sentimentsparsity} mitigates the sparsity and irregularity of tweets with an additional feature set by extracting latent topics and the associated topic sentiment from tweets. \cite{offensivetweets} used a semi-supervised topic model to detect offensive tweets, showing that topic modeling could outperform a keyword matching approach. While topical features are strong, we argue that additional geographic information can significantly improve results. \cite{geoloc} estimates city-level location using tweet content, finding 51 percent of Twitter users within 100 miles of their actual location. With their work and others in mind, we choose to leverage geographic location as a feature of a text corpus.\newline
\indent Our study builds on previous work in STI prediction by constructing a multi-dimensional feature set. We hypothesize that counties in the same geographic region have similar topical distribution, so for a given county, neighboring counties would be a good reflection of that county. We show that multi-dimensional feature construction can significantly improve the STI predictive power of topic features alone with respect to tweets.

\section{Dataset and Task Definition}

We obtained zip code-level HIV diagnosis rates per 100,000 from Philadelphia, Pennsylvania which the HIV data included only people aged 13 and older. Data from regions with less than 5 new HIV diagnoses per year or less than 100 inhabitants are routinely suppressed by the CDC, and this suppression criteria were also applicable for the present analysis.\newline
\indent Our Twitter corpus ranges from June 2009 to March 2010, November 2011 to December
2015 with more than 3.4 billion tweets, including re-tweets. However in order to use this dataset at the spacial granularity of the STI new diagnosis rates we geotagged our Twitter corpus to zipcodes, and counties, in the US.
We choose a training set (Twitter corpus 2009-2014) to learn a predictive LDA model\cite{lda}. Our baseline predictive model is based on topic features generated by a tf-idf bag-of-words representation of the corpus. The ability of these features to capture indicators of risky behavior will determine the accuracy of our HIV prediction.For the prediction task, we experimented with Multinomial Naive Bayes, Support Vector Machines, and other classifiers. In this paper, we focus on building multi-dimensional features using location information and instances of slang. 
\section{Multi-dimensional Feature Construction}

\subsection{Regional Colloquialism Features}
Our goal here was to find topical likeness between tweets that used the same slang terms. We define slang as words in online slang dictionaries, similar to \cite{generaltopicmodelling}. We stripped each tweet of words that could not be classified as slang in order to minimize overlap between disparate and consequently dissimilar counties. This approach is novel in detecting geographic location in local diction. We hypothesized that these exclusively slang tweets would better capture local colloquialisms. To avoid sparse feature set (since many tweets contained no slang words), we created an additional feature set to represent the number of slang words relative to the length of the tweet in order to preserve more tweet content.

\subsection{Smoothing-based Features}
As discussed previously, our tweets were distributed into documents based on county\cite{geoloc}. For tweets without available location information, we generated a topic model for each document (represents what tweets we already have for a given county), and measured topical similarity of each tweet to the topic distribution of each county. \newline
\indent For many counties, minimal tweets produced inaccurate and skewed topics. To counter this, we used neighboring counties to smooth the topic distributions of each county. We assume that neighboring counties share similar topical distribution, so the feature set produced by neighboring counties in combination with a county's features should be more reliable.
\indent For the concatenated smoothing method, we appended the additional topics to our baseline. We add another feature set for average topic distribution of neighboring zip codes, resulting double the original topics. For both methods, we experimented with radius of neighboring zip codes, and \textit{multiplier} (which we use to offset the weight of a neighboring topic distribution).
\newline
\begin{figure}[h]

\includegraphics[width=80mm,scale=0.5]{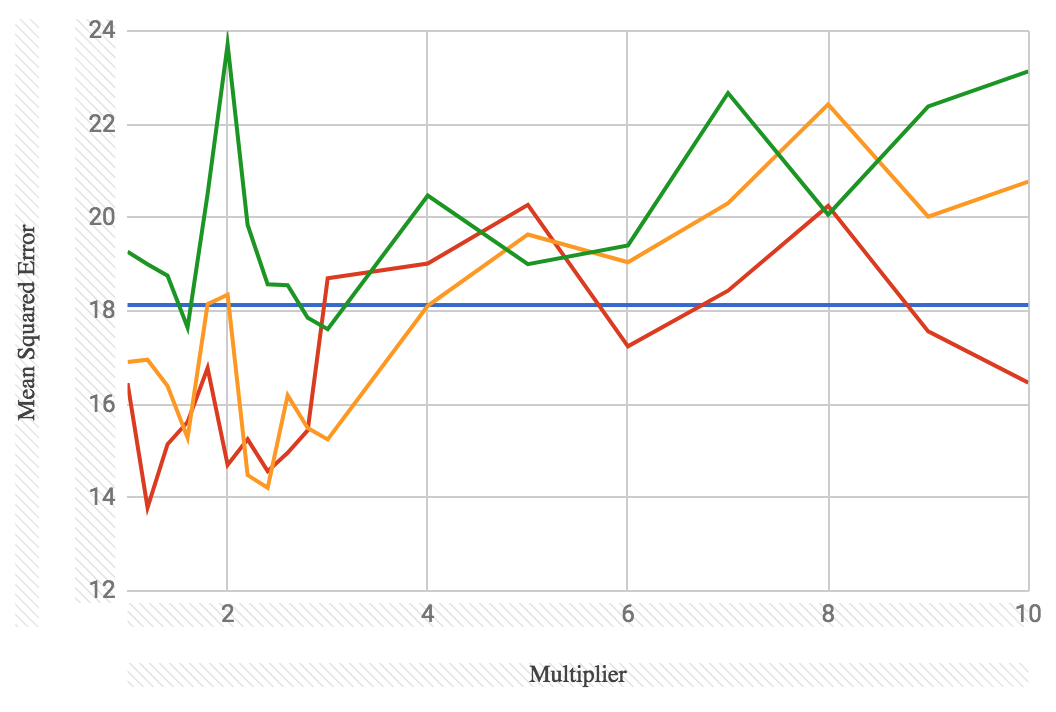}
\caption{MSE for varying weights for neighboring distributions: {\textcolor{blue}{Baseline}}, {\textcolor{red}{Smoothing-based}}, {\textcolor{ForestGreen}{Slang-based}}, {\textcolor{Orange}{Smoothing and Slang-based}}, where the multiplier represents the relative weight of neighboring topic distributions.}
\end{figure}

\subsection{Mortality Rates Experiments}
\subsubsection{Dataset and Task Definition}
We obtained the crude rate for opioid usage, suicide, and mental disorder at the FIPS code level across the country for 2014 - 2016. Accordingly, our Twitter corpus ranges from January 2014 - December 2016. 
We exclusively used crude rate data for which we had tweets and geo-located tweets for which we had crute rates. 
Our task is to use a Twitter corpus to predict crude rates through classification of feature sets.
We used feature sets generated from 2014-2015 crude rates to train 2016 to test.
\\\\
\subsubsection{Feature Set}
In order to construct a feature set, we first generated a series of LDA topic models from the Twitter corpus. The number of topics ranged from $5-200$. Then, we applied each $n$-topic model to each document (set of tweets) in the Twitter corpus, to produce $n$ features for each document. Each feature maps to a topic, representing the topic distribution of the document. We use these features to predict the crude rates for opioid usage, suicide, and mental disorders, assuming that this behavior is reflected somehow in the topic distribution of user's tweets.
\subsubsection{Classification}

Once we have a feature set representing the training (2014-2015 tweets) and testing (2016 tweets) corpus, we can use classification to predict crude rates. 
We lumped each crude rate into one of 6 labels depending on standard deviation from the mean.
We trained the following classifiers:
\begin{itemize}
    \item Bernoulli naive Bayes
    \item Gaussian naive Bayes
    \item Multinomial naive Bayes
    \item K-Neighbors
    \item Random Forest
\end{itemize}
Since there were six labels, our baseline accuracy for random guess is $.16$. %Our results are depicted below.

\subsection{Results and Conclusion}

With respect to smoothing-based features and location-based features in combination with colloquialism features, our experiments surpassed our baseline.\newline  
\indent We observed that adding the slang-term feature set was less successful than the smoothing-based feature set. When we reduced the corpus to solely tweets containing slang terms, it is likely that this additional feature set, equally weighted with the original, caused the slang feature set to have too much weight. While the tweets in the slang-corpus better capture the regional colloquialisms, we are still reducing the the number of examples our topic model has available to classify in the training set.\newline
\indent Overall, adding the smoothing-based feature set significantly improved our baseline. It is likely, that such multi-dimensional feature construction is an effective way improve a set of topic features. In the future, more exploration in alternative ways to represent geographic location in textual data is necessary to take full advantage of multi-dimensional features. 
\subsection*{Acknowledgement}

We would like to thank Kathleen A. Brady, from the
Philadelphia department of public health, for providing
us with the Philadelphia HIV new diagnosis datasets.
This work was supported by the National Institutes of
Health, Grant 1 R56 AI114501-01A1. We would also like to thank Professor ChengXiang Zhai for all of his help and guidance in making this poster submission possible.
\\
%\subsection{References}

\end{document}